\documentclass[12pt]{article}
\usepackage{a4wide}
\usepackage{epsfig}
\usepackage{amsmath}

\newlength{\absize}
\catcode`@=11
\def\citer{\@ifnextchar [{\@tempswatrue\@citexr}{\@tempswafalse\@citexr[]}}

\def\@citexr[#1]#2{\if@filesw\immediate
  \write\@auxout{\string\citation{#2}}\fi
  \def\@citea{}\@cite{\@for\@citeb:=#2\do
    {\@citea\def\@citea{--\penalty\@m}\@ifundefined
       {b@\@citeb}{{\bf ?}\@warning
       {Citation `\@citeb' on page \thepage \space undefined}}%
\hbox{\csname b@\@citeb\endcsname}}}{#1}}
\catcode`@=12

\begin{document}
  \thispagestyle{empty}
  \pagestyle{empty}
  \renewcommand{\thefootnote}{\fnsymbol{footnote}}
\newpage\normalsize
    \pagestyle{plain}
    \setlength{\baselineskip}{4ex}\par
    \setcounter{footnote}{0}
    \renewcommand{\thefootnote}{\arabic{footnote}}
\newcommand{\preprint}[1]{%
  \begin{flushright}
    \setlength{\baselineskip}{3ex} #1
  \end{flushright}}
\renewcommand{\title}[1]{%
  \begin{center}
    \LARGE #1
  \end{center}\par}
\renewcommand{\author}[1]{%
  \vspace{2ex}
  {\Large
   \begin{center}
     \setlength{\baselineskip}{3ex} #1 \par
   \end{center}}}
\renewcommand{\thanks}[1]{\footnote{#1}}
\vskip 0.5cm

\begin{center}
{\large \bf Induced Fractional Zero-Point Angular Momentum for
Charged Particles of the Bohm-Aharonov System by means of a
``Spectator" Magnetic Field$\;^ {\ast}$}
\end{center}
\vspace{1cm}
\begin{center}
Jian-Zu Zhang
\end{center}
\begin{center}
Institute for Theoretical Physics, East China University of
Science and Technology, Box 316, Shanghai 200237, P. R. China
\end{center}
\vspace{1cm}

\begin{abstract}
An induced fractional zero-point angular momentum of charged
particles by the Bohm-Aharonov (B-A) vector potential is realized
via a modified combined trap.  It explores a ``spectator"
mechanism in this type of quantum effects: In the limit of the
kinetic energy approaching one of its eigenvalues the B-A vector
potential alone cannot induce a fractional zero-point angular
momentum at quantum mechanical level in the B-A magnetic
field-free region; But when there is a ``spectator" magnetic field
the B-A vector potential induces a fractional zero-point angular
momentum. The ``spectator" does not contribute to such a
fractional angular momentum, but plays essential role in
guaranteeing non-trivial dynamics at quantum mechanical level in
the required limit. This ``spectator" mechanism is significant in
investigating the B-A effects and related topics in both aspects
of theory and experiment.
\end{abstract}

\begin{flushleft}
$^{\ast}$ Revised version of arXiv: 0709.0843
\end{flushleft}
\clearpage
As is well known, quantum states of charged particles can be
influenced by electromagnetic effects even if those particles are
in a region of vanishing field strength \cite{ES,AB}. As predicted
by Bohm and Aharonov (B-A) \cite{AB},
experiments \cite{expt} showed that in a multiply connected region
where field strength is zero everywhere the interference spectrum
suffered a shift according to the amount
of the loop integral of magnetic vector potential around an
unshrinkable loop. Wu and Yang \cite{WY} pointed out that the B-A
effects is due to the non-trivial topology of the space wehere the
magnetic field strength is vanishing.
The B-A effect is purely quantum mechanical one which explores
far-reaching consequences of vector potential in quantum theory.
This effect has been received much attention for years
\cite{OP,PT89,M-K04}. Recently investigations in this topic
concentrated on revealing new types of quantum phases: The
Aharonov-Casher effect \cite{AC}, the He-McKellar-Wilkens phase
\cite{HMW} and the Anandan phase \cite{Anan}.

\vspace{0.1cm}

In another aspect a fractional
angular momentum originated from the Poynting vector produced by
crossing the Coulomb field of a charged particle with an external
magnetic field has been predicted by Peshkin, Talmi and Tassie for
years \cite{PT89,PT60}.
There are lots of works concerning fractional angular momentum in
B-A dynamics and their ¡°fractional¡± statistics (see the reviews
\citer{Wilc,Laug} and references therein). Spatial
noncommutativity also leads to fractional angular momentum
\cite{JZZ04a,JZZ04b}.

\vspace{0.1cm}

Recently Kastrup \cite{Kast06} considered the question of how to
quantize a classical system of the canonically conjugate pair
angle and orbital angular momentum. This has been a controversial
issue since the founding days of quantum mechanics \cite{Kast03}.
The problem is that the angle is a multivalued or discontinuous
variable on the corresponding phase space. A crucial point is that
the irreducible unitary representations of the euclidean group
$E(2)$ or of its covering groups allow for orbital angular
momentum $l = \hbar(n+\delta)$ where $n=0,\pm 1,\pm 2,\cdots,$ and
$0\le \delta<1$. The case $\delta\ne 0$ corresponds to fractional
zero-point angular momentum. Kastrup investigated the physical
possibility of fractional orbital angular momentum in connection
with the quantum optics of Laguerre-Gaussian laser modes in
external magnetic fields, and pointed out that if implementable
this would lead to a wealth of new theoretical, experimental and
even technological possibilities.

\vspace{0.1cm}

In this paper the induced fractional zero-point angular momentum
of charged particles by the B-A vector potential is realized via a
modified combined trap.  It explores a ``spectator" mechanism in
this type of quantum effects: In the limit of the kinetic energy
approaching one of its eigenvalues the B-A vector potential alone
cannot induce a fractional zero-point angular momentum of charged
particles at quantum mechanical level in a region of vanishing B-A
field strength; But when there is a ``spectator" magnetic field
the B-A vector potential induces a fractional zero-point angular
momentum in the same region. The ``spectator" does not contribute
to such a fractional angular momentum, but plays essential role in
guaranteeing non-trivial dynamics at quantum mechanical level in
the required limit.
This type of quantum effects is so remarkable that in quantum
mechanics the vector potential itself has physical significant
meaning and becomes effectively measurable not only in shifts of
interference spectra originated from quantum phases but also in
physical observables.

\vspace{0.1cm}

{\bf 1. Dynamics in a Modified Combined Trap} -- We consider ions
constrained in a modified combined trap including the B-A type
magnetic field. The Paul, Penning, and combined traps share the
same electrode structure \cite{BG}.
A combined trap operates in all of the fields of the
Paul and Penning traps being applied simultaneously. The trapping
mechanism in a Paul trap involves an oscillating axially symmetric
electric potential
$\tilde{U}(\rho,\phi,z,t)= U(\rho,\phi,z)cos\tilde{\Omega} t$
with
$U(\rho,\phi,z)=V(z^2-\rho^2/2)/2d^2$
where $\rho$,
$\phi$ and $z$ are cylindrical
coordinates, $V$ and $d$ are, respectively, characteristic voltage
and length, and $\tilde{\Omega}$ is a large radio-frequency. The
dominant effect of the oscillating potential is to add an
oscillating phase factor to the wave function. Rapidly varying
terms of time in Schro¡§dinger equation can be replaced by their
average values. Thus for
$\tilde{\Omega}\gg \Omega\equiv \left(\sqrt{2}q |V|/\mu d^2
\right)^{1/2}$
we obtain a time-independent effective electric potential
\cite{CSW}
$V_{eff}=q^2\nabla U\cdot\nabla U/4\mu\tilde{\Omega}^2=
\mu\omega_P^2(\rho^2+4z^2)/2$
where $\mu$ and $q(>0)$ are, respectively, the mass and charge of
the trapped
ion, and
$\omega_P=\Omega^2/4\tilde{\Omega}$.
A modified combined trap combines the above electrostatic
potential and two magnetic fields \cite{note-1}: a homogeneous
magnetic field ${\bf B}_c$ aligned along the $z$ axis in a normal
combined trap and a B-A type magnetic field ${\bf B}_0$ produced
by, for example, an infinitely long solenoid with radius
$\rho=(x_1^2+x_2^2)^{1/2}=a$. Inside the solenoid $(\rho<a)\;$
${\bf B}_{0,in}=(0, 0, B_0)$ is homogeneous along the $z$ axis,
and outside the solenoid $(\rho>a)$ ${\bf B}_{0,out}=0$. The
vector potential ${\bf A}_c$ of ${\bf B}_c$ is chosen as
(Henceforth the summation convention is used)
$A_{c,i}=-B_c\epsilon_{ij}x_j/2$, $A_{c,z}=0,\; (i,j=1,2).$
The B-A vector potential ${\bf A}_0$ is: Inside the solenoid
$A_{0,i}=A_{in,i}=-B_0\epsilon_{ij}x_j/2$,\; $A_{in,z}=0;$
Outside the solenoid
$A_{0,i}=A_{out,i}=-B_0 a^2 \epsilon_{ij}x_j/2x_k
x_k,\;A_{out,z}=0,\; (i,j,k=1,2).$
At $\rho=a$ the potential ${\bf A}_{in}$ passes continuously over
into ${\bf A}_{out}$.
The Hamiltonian of the modified combined
trap is
$H=\left(p_i-qA_{c,i}/c-qA_{0,i}/c\right)^2/2\mu +p_z^2/2\mu+
\mu\omega_P^2\left(x_i^2+4z^2\right)/2$.
This Hamiltonian can be decomposed into a one-dimensional harmonic
Hamiltonian $H_z(z)$ along the $z$-axis with the axial frequency
$\omega_z=2\omega_P$ and a two-dimensional Hamiltonian
$H_{\perp}(x_1,x_2)$, $H=H_z(z)+H_{\perp}(x_1,x_2)$. Inside the
solenoid the ion's motion is the same as the one with a total
magnetic field ${\bf B}_c+{\bf B}_{0,in}$.

\vspace{0.1cm}

In the following we consider the motion outside the solenoid. The
two-dimensional Hamiltonian outside the solenoid is \cite{BG,CSW}
\begin{equation}
\label{Eq:H1}
H_{\perp}(x_1,x_2)=\frac{1}{2\mu}\left(p_i+\frac{1}{2}\mu\omega_c
\epsilon_{ij}x_j+\mu\omega_0 a^2\frac{\epsilon_{ij}x_j}{2x_k
x_k}\right)^2+ \frac{1}{2}\mu\omega_P^2 x_i^2,
\end{equation}
where $\omega_c=qB_c/\mu c$ and $\omega_0=qB_0/\mu c$ are the
cyclotron frequencies corresponding to, respectively, the magnetic
fields ${\bf B} _c$ and ${\bf B}_{0,in}$. The Hamiltonian
$H_{\perp}$ possess a rotational symmetry in $(x_1, x_2)$ - plane.
The $z$-component of the orbital angular momentum
$J_z=\epsilon_{ij} x_i p_j$
commutes with $H_{\perp}$. They have common eigenstates.

\vspace{0.1cm}

{\it Dynamics in the Limit of the Kinetic Energy Approaching its
Lowest Eigenvalue} -- In this limit the kinetic energy is
$E_k=\mu \dot{x_i} \dot{x_i}/2=\left(K_1^2+K_2^2\right)/2\mu$
where
\begin{equation}
\label{Eq:K1}
K_i\equiv p_i+\frac{1}{2}\mu\omega_c \epsilon_{ij}x_j +\mu\omega_0
a^2\frac{\epsilon_{ij}x_j}{2x_k x_k},\quad
[K_i,K_j]=i\hbar\mu\omega_c\epsilon_{ij}
\end{equation}
Here $K_i$ is the mechanical momenta corresponding to the vector
potentials $A_{c,i}$ and $A_{out,i}$. It is worth noting that the
B-A vector potential $A_{out,i}$ does {\it not} contributes to the
commutator $[K_i,K_j]$. The canonical momenta $p_i$ are quantized,
$p_i=-i\hbar\partial/\partial x_i$. They commute each other
$[p_i,p_j]=0.$
We define canonical variables $Q=K_1/\mu\omega_c$ and $\Pi=K_2$
which satisfy
$[Q,\Pi]=i\hbar\delta_{ij}.$
The kinetic energy $E_k$ is rewritten as the Hamiltonian of a
harmonic oscillator
$E_k=\Pi^2/2\mu+\mu\omega_c^2 Q^2/2.$
The lowest eigenvalue $\mathcal{E}_{k0}$ of the kinetic energy
$E_k$ is \cite{private}
$\mathcal{E}_{k0}=\hbar\omega_c/2.$

\vspace{0.1cm}

In a laser trapping field, using a number of laser beams and
exploiting Zeeman tuning, the speed of atoms can be slowed to the
extent of $1\;ms^{-1}$, see \cite{note-2}.
Ions are the common object in cooling and trapping. In order to
experimentally realizing the limit of $E_k\to\mathcal{E}_{k0}$
through laser cooling in a trap ions are used.

\vspace{0.1cm}

In the limit of the kinetic energy approaching its lowest
eigenvalue the Hamiltonian $H_{\perp}$ in Eq.~(\ref{Eq:H1}) has
non-trivial dynamics \cite{Baxt,JZZ96,JZZ04b}. The Lagrangian
corresponding to $H_{\perp}$ is
\begin{equation}
\label{Eq:L1}
L=\frac{1}{2}\mu\dot{x_i}\dot{x_i} -\frac{1}{2}\mu\omega_c
\epsilon_{ij}\dot{x_i}x_j-\mu\omega_0
a^2\frac{\epsilon_{ij}\dot{x_i}x_j}{2x_k
x_k}-\frac{1}{2}\mu\omega_P^2 x_i x_i.
\end{equation}
In the limit of $E_k\to\mathcal{E}_{k0},$
the Hamiltonian $H_{\perp}$ reduces to
$H_0=\hbar\omega_c/2 + \mu\omega_P^2 x_i x_i/2.$
The Lagrangian corresponds to $H_0$ is
\begin{equation}
\label{Eq:L2}
L_0=-\frac{1}{2}\mu\omega_c \epsilon_{ij}\dot{x_i}x_j-\mu\omega_0
a^2\frac{\epsilon_{ij}\dot{x_i}x_j}{2x_k x_k}
-\frac{1}{2}\mu\omega_P^2 x_i x_i-\frac{1}{2}\hbar\omega_c.
\end{equation}

\vspace{0.1cm}

{\it Constraints} -- For the reduced system $(H_0,L_0)$ the
canonical momenta are
\begin{equation}
\label{Eq:p1}
p_{i}=\frac{\partial L_0}{\partial
\dot{x_i}}=-\frac{1}{2}\mu\omega_c \epsilon_{ij}x_j-\mu\omega_0
a^2\frac{\epsilon_{ij}x_j}{2x_k x_k}.
\end{equation}
Eq.~(\ref{Eq:p1}) does not determine velocities $\dot{x_i}$ as
functions of $p_{i}$ and $x_j$, but gives relations among $p_{i}$
and $x_j$, that is, such relations are the primary constraints
\cite{M-K06,JZZ96,JZZ04b}
\begin{equation}
\label{Eq:C1}
\varphi_i(x,p)=p_i +\frac{1}{2}\mu\omega_c
\epsilon_{ij}x_j+\mu\omega_0 a^2\frac{\epsilon_{ij}x_j}{2x_k
x_k}=0.
\end{equation}
The physical meaning of Eq.~(\ref{Eq:C1}) is that it expresses the
dependence of degrees of freedom among $p_{i}$ and $x_j$.
The constraints (\ref{Eq:C1}) should be carefully treated
\cite{constraint}. The subject can be treated simply by the
symplectic method in \cite{FJ,DJ93}. In this paper we work in the
Dirac formalism. The Poisson brackets of the constraints
(\ref{Eq:C1}) are
\begin{equation}
\label{Eq:Poisson-1}
C_{ij}=\{\varphi_i, \varphi_j\}= \mu\omega_c\epsilon_{ij}.
\end{equation}
From Eq.~(\ref{Eq:Poisson-1}), $\{\varphi_i, \varphi_j\}\ne 0,$ it
follows that the conditions of the constraints $\varphi_i$ holding
at all times do not lead to secondary constraints.

\vspace{0.1cm}

$C_{ij}$ defined in Eq.~(\ref{Eq:Poisson-1}) are elements of the
constraint matrix $\mathcal{C}.$ Elements of its inverse matrix
$\mathcal{C}^{-1}$ are $(C^{-1})_{ij}=-\epsilon_{ij}/\mu\omega_c.$
The corresponding Dirac brackets of $\{\varphi_i, x_j\}_D$,
$\{\varphi_i, p_j\}_D$, $\{x_i, x_j\}_D$, $\{p_i, p_j\}_D$ and
$\{x_i, p_j\}_D$ can be defined. The Dirac brackets of $\varphi_i$
with any variables $x_i$ and $p_j$ are zero so that the
constraints (\ref{Eq:C1}) are strong conditions. It can be used to
eliminate dependent variables. If we select $x_1$ and $x_2$ as the
independent variables, from the constraints (\ref{Eq:C1}) the
variables $p_1$ and $p_2$ can be represented by, respectively, the
independent variables $x_2$ and $x_1$ as
\begin{equation}
\label{Eq:p-x-1}
p_1=-\frac{1}{2}\mu\omega_c x_2-\mu\omega_0 a^2\frac{x_2}{2x_k
x_k}, \;
p_2=\frac{1}{2}\mu\omega_c x_1+\mu\omega_0 a^2\frac{x_1}{2x_k x_k}
\end{equation}
The Dirac brackets of $x_1$ and $x_2$ is
\begin{equation}
\label{Eq:Dirac}
\{x_1,x_2\}_D=\frac{1}{\mu\omega_c}.
\end{equation}
We introduce new canonical variables $x=x_1$  and $p=\mu\omega_c
x_2.$ Their Dirac bracket is $\{x,p\}_D=1.$ According to Dirac's
formalism of quantizing constrained systems the corresponding
quantum commutation relation is $[x,p]=i\hbar$.

\vspace{0.1cm}

{\it Quantum Behavior of the Reduced System} -- Now we consider
quantum behavior of the reduced system $(H_0, L_0)$. By defining
the following effective mass and frequency,
$\mu^{\ast}\equiv \mu\omega_c^2/\omega_P^2, \quad \omega^{\ast}
\equiv\omega_P^2/\omega_c,$
the Hamiltonian $H_0$ is represented as
$H_0=p^2/2\mu^{\ast}+\mu^{\ast}{\omega^{\ast}}^2x^2/2+\hbar\omega_c/2.$
We introduce an annihilation operator
$A= \sqrt{\mu^{\ast}\omega^{\ast}/2\hbar}\;x
+i\sqrt{1/2\hbar\mu^{\ast}\omega^{\ast}}\;p$
and its conjugate one $A^\dagger.$ The operators $A$ and
$A^\dagger$ satisfies $[A,A^\dagger]=1$. The eigenvalues of the
number operator $N=A^\dagger A$ is $n=0, 1, 2, \cdots$. Using $A$
and $A^\dagger$, the reduced Hamiltonian $H_0$ is rewritten as
$H_0=\hbar\omega^{\ast}\left(A^\dagger A + 1/2\right)+
\hbar\omega_c/2.$

\vspace{0.1cm}

Now we consider the angular momentum of the ion. Using
Eq.~(\ref{Eq:p-x-1}) to replace $p_1$ and $p_2$ by, respectively,
the independent variables $x_2$ and $x_1$, the orbital angular
momentum $J_z=\epsilon_{ij}x_i p_j$ is rewritten as
\begin{equation}
\label{Eq:J}
J_z=\frac{q}{2\pi c}\Phi_0 + \frac{1}{2}\mu\omega_c(x_1^2+x_2^2),
\end{equation}
where $\Phi_0=\pi a^2 B_0$ is the total flux of the magnetic field
$B_0$ inside the solenoid. Similarly, using $A$ and $A^\dagger$ to
rewrite $J_z$, we obtain
$J_z=q\Phi_0/2\pi c + \hbar\left(A^\dagger A+1/2\right).$
The zero-point angular momentum of $J_z$ is
$\mathcal{J}_0=\hbar/2+q\Phi_0/2\pi c.$ In the above the term
\cite{note-3}
\begin{equation}
\label{Eq:J-AB}
\mathcal{J}_{AB}=\frac{q}{2\pi c}\Phi_0
\end{equation}
is the zero-point angular momentum induced by the AB vector
potential. $\mathcal{J}_{AB}$ takes fractional values. It is
related to the region where the magnetic field ${\bf B}_{0,out}=0$
but the corresponding vector potential ${\bf A}_{out} \ne 0.$

\vspace{0.1cm}

{\bf 2. Dynamics in the Case of $\bf {B_c=0}$} -- It is worth
noting that here ${\bf B}_c$, like a ``spectator", does not
contribute to $\mathcal{J}_{AB}$. In order to clarify the role
played by ${\bf B}_c$, we consider the case of ${\bf B}_c=0$. In
this case the modified combined trap is as stable as a Paul trap.
The corresponding kinetic energy reduces to
$\tilde E_k=\mu \dot{x_i} \dot{x_i}/2=\left(\tilde K_1^2+\tilde
K_2^2\right)/2\mu$
where
\begin{equation}
\label{Eq:K2}
\tilde K_i\equiv p_i+\mu\omega_0 a^2\frac{\epsilon_{ij}x_j}{2x_k
x_k}, \quad
[\tilde K_i,\tilde K_j]=0.
\end{equation}
In the above $\tilde K_i$ is the mechanical momenta corresponding
to the B-A vector potential $A_{out,i}$. Unlike the ordinary
vector potential, the special feature of the B-A vector potential
is that it does {\it not} contributes to the commutator $[\tilde
K_i,\tilde K_j]$. Because $\tilde K_i$ are commuting, behavior of
$\tilde E_k$ is similar to a Hamiltonian of a free particle. Its
spectrum is a continuous one. When $\tilde E_k$ approaching some
constant $\mathcal{\tilde E}_k(\ne 0)$ the Hamiltonian $H_{\perp}$
reduces to
$\tilde H_0=\mathcal{\tilde E}_k+\mu\omega_P^2 x_i x_i/2.$
The Lagrangian corresponding to $\tilde H_0$ is
\begin{equation}
\label{Eq:L3}
\tilde L_0=-\mu\omega_0 a^2\frac{\epsilon_{ij}\dot{x_i}x_j}{2x_k
x_k} -\frac{1}{2}\mu\omega_P^2 x_i x_i-\mathcal{\tilde E}_k.
\end{equation}
From $\tilde L_0$ we obtain the canonical momenta
\begin{equation}
\label{Eq:p2}
\tilde p_{i}=\frac{\partial\tilde {L_0}}{\partial \dot{x_i}}
=-\mu\omega_0 a^2\frac{\epsilon_{ij}x_j}{2x_k x_k}.
\end{equation}
Now we clarify that the case $\mathcal{\tilde E}_k=0$ should be
excluded. The limit of the kinetic energy $E_k=\mu \dot{x_i}
\dot{x_i}/2\to 0$ corresponds two possibilities: $\dot{x_i}=0$ or
$\mu\to 0.$ In the case $\dot{x_i}=0$ the Lagrangian $L$ in
Eq.~(\ref{Eq:L1}) reduces to $\tilde L_0^{\prime}=-\mu\omega_P^2
x_i x_i/2.$ The corresponding canonical momenta $\tilde
p_{i}=\partial \tilde L_0^{\prime}/\partial \dot{x_i}=0.$
Therefore there is no dynamics. According to the definition of the
frequency $\Omega$ the other possibility $\mu\to 0$ is forbidden.

\vspace{0.1cm}

Eq.~(\ref{Eq:p2}) gives the reduced primary constraints
\begin{equation}
\label{Eq:C2}
\tilde \varphi_i=\tilde p_{0i}+\mu\omega_0
a^2\frac{\epsilon_{ij}x_j}{2x_k x_k}=0.
\end{equation}
Here the special feature is that the corresponding Poisson
brackets are zero,
\begin{equation}
\label{Eq:Poisson-2}
\tilde C_{ij}=\{\tilde{\varphi_i}, \tilde{\varphi_j}\}\equiv 0.
\end{equation}
From Eq.~(\ref{Eq:Poisson-2}), $\{\tilde \varphi_i, \tilde
\varphi_i\}\equiv 0,$ it follows that the conditions of the
constraints $\tilde\varphi_i$ holding at all times lead to
secondary constraints
$\tilde\varphi_i^{(2)}=-\mu\omega_P^2 x_i$.
The Poisson brackets
$\{\tilde{\varphi_i}^{(2)}, \tilde{\varphi_j}\}=0$,
$\{\tilde{\varphi_i}^{(2)}, \tilde{\varphi_j}^{(2)}\}=0$, and
$\{\tilde{\varphi_i}^{(2)}, \tilde H_0\}=0$,
so that persistence of the secondary constraints
$\tilde\varphi_i^{(2)}$ in course of time does not lead to further
secondary constraints $\tilde\varphi_i^{(3)}$.

\vspace{0.1cm}

Because of $\tilde C_{ij}\equiv 0,$ the inverse matrix
$\mathcal{\tilde C}^{-1}$ does not exist. The Dirac brackets
$\{\tilde{\varphi_i}, x_j\}_D$, $\{\tilde{\varphi_i}, p_j\}_D$,
$\{\tilde{\varphi_i}^{(2)}, x_j\}_D$, $\{\tilde{\varphi_i}^{(2)},
p_j\}_D$, $\{x_i, x_j\}_D$, $\{p_i, p_j\}_D$, and $\{x_i, p_j\}_D$
cannot be defined. According to Dirac's formalism of quantizing
constrained systems, there is no way to establish dynamics at
quantum mechanical level. This means that the B-A vector potential
alone cannot lead to non-trivial dynamics at quantum mechanical
level in the required limit, thus does {\it not} contribute to the
energy spectrum and angular momentum at all.

It is clear that though the vector potential $A_{c,i}$ of the
``spectator" magnetic field ${\bf B}_c$ does not contribute to
$\mathcal{J}_{AB}$, it plays essential role in guaranteeing
non-trivial dynamics at quantum mechanical level in the limit of
the kinetic energy approaching one of its eigenvalues.
This example reveals that, unlike ordinary vector potential, the
physical role played by the B-A vector potential is subtle. This
needs to be carefully analyzed at quantum mechanical level.

\vspace{0.1cm}

{\bf 3. Dynamics in the Case of $\bf {B_0=0}$} -- In order to
further clarify the essential difference between ${\bf A}_o$ and
${\bf A}_c$ in the region of ${\bf B}_{0,out}=0$ we consider the
case of ${\bf B}_0=0.$ In this case the modified combined trap
reduces to a combined trap. The Hamiltonian $H_{\perp}(x_1,x_2)$
in Eq.~(\ref{Eq:H1}) reduces to
$\hat H_{\perp}(x_1,x_2)=\left(p_i+\mu\omega_c
\epsilon_{ij}x_j/2\right)^2/2\mu + \mu\omega_P^2 x_i^2/2.$
Its kinetic energy is
$\hat E_k=(\hat K_1^2+\hat K_2^2)/2\mu$
where
\begin{equation}
\label{Eq:K3}
\hat K_i\equiv p_i+\mu\omega_c \epsilon_{ij}x_j/2, \quad
[\hat K_i,\hat K_j]=i\hbar\mu\omega_c\epsilon_{ij}.
\end{equation}
In Eq.~(\ref{Eq:K3}) $\hat K_i$ is the mechanical momenta
corresponding to the vector potentials $A_{c,i}$. The commutation
relations between $\hat K_i$'s are the same as the ones between
$K_i$'s in Eq.~(\ref{Eq:K1}). The eigenvalues of $\hat E_k$ is
$\mathcal{\hat E}_{kn}=\hbar\omega_c(n+1/2),$
which are just the Landau levels of charged particles in an
external magnetic field.

\vspace{0.1cm}

In the following we consider the limit of $\hat E_k$ approaching
the lowest eigenvalue $\mathcal{\hat E}_{k0}=\hbar\omega_c/2.$
The Lagrangian corresponding to $\hat H_{\perp}$ is
\begin{equation}
\label{Eq:L4}
\hat L=\mu\dot{x_i}\dot{x_i}/2 - \mu\omega_c
\epsilon_{ij}\dot{x_i}x_j/2 - \mu\omega_P^2 x_i x_i/2.
\end{equation}
In the limit of $\hat E_k\to\mathcal{\hat E}_{k0},$
the Hamiltonian $\hat H_{\perp}$ reduces to
$\hat H_0=\hbar\omega_c/2+\mu\omega_P^2 x_i x_i/2$
which is the same as $H_0$. The Lagrangian  corresponds to $\hat
H_0$ is
\begin{equation}
\label{Eq:L5}
\hat L_0=-\mu\omega_c \epsilon_{ij}\dot{x_i}x_j/2 -\mu\omega_P^2
x_i x_i/2-\hbar\omega_c/2.
\end{equation}
For the reduced system $(\hat H_0,\hat L_0)$ the canonical momenta
are
$\hat p_{i}=\partial \hat L_0/\partial \dot{x_i}=-\mu\omega_c
\epsilon_{ij}x_j/2.$
It leads to the following constraints
\begin{equation}
\label{Eq:C3}
\hat \varphi_i=p_i + \mu\omega_c \epsilon_{ij}x_j/2=0.
\end{equation}
The Poisson brackets of $\hat \varphi_i$ are the same as ones of
the constraints $\varphi_i$ in Eq.~(\ref{Eq:Poisson-1}):
\begin{equation}
\label{Eq:Poisson-3}
\hat C_{ij}=\{\hat \varphi_i, \hat \varphi_i\}=
\mu\omega_c\epsilon_{ij}.
\end{equation}
From Eq.~(\ref{Eq:Poisson-3}), $\{\hat \varphi_i, \hat
\varphi_i\}\ne 0,$ it follows that the conditions of the
constraints $\hat \varphi_i$ holding at all times do not lead to
secondary constraints.

\vspace{0.1cm}

By the similar procedure of treating the constraints
(\ref{Eq:C1}), we find that the reduced system $(\hat H_0,\hat
L_0)$ has non-trivial dynamics at quantum mechanical level in the
limit of $\hat E_k\to\mathcal{\hat E}_{k0}.$ The constraints
(\ref{Eq:C3}) are strong conditions  which can be used to
eliminate dependent variables. We select $x_1$ and $x_2$ as the
independent variables. The variables $p_1$ and $p_2$ can be
represented by, respectively, $x_2$ and $x_1$ as
$p_1=-\mu\omega_c x_2/2, \; p_2=\mu\omega_c x_1/2.$
The Dirac brackets of $x_1$ and $x_2$ is
$\{x_1,x_2\}_D=1/\mu\omega_c.$
We introduce new canonical variables $x=x_1$  and $p=\mu\omega_c
x_2.$ Their Dirac bracket is $\{x,p\}_D=1.$ The corresponding
quantum commutation relation is $[x,p]=i\hbar$. Using these
results the orbital angular momentum $J_z=\epsilon_{ij}x_i p_j$
can be represented by the canonical variables $x$ and $p$ as
$\hat J_z=\left(p^2/2\mu+\mu\omega_c^2 x^2/2\right)/\omega_c.$
The zero-point angular momentum can be read out from this
harmonic-like ``Hamiltonian",
$\mathcal{\hat J}_0=\hbar/2.$
We note that in this case there is {\it no} fractional zero-point
angular momentum.

\vspace{0.1cm}

The above results elucidate that ${\bf A}_c$ are essentially
different from ${\bf A}_0$: the ${\bf A}_c$ alone can lead to
non-trivial dynamics at quantum mechanical level in the limit of
the kinetic energy approaching its lowest eigenvalue.

\vspace{0.1cm}

{\bf 4. Gauge Transformation} -- As is well known, we can perform
a gauge transformation $\chi$ so that the resulting vector
potential
${\bf A}_{out}^{\prime}={\bf A}_{out}+\nabla \chi =0.$
A suitable gauge function \cite{note-4} is
$\chi =-B_0 a^2 tan^{-1}(x_2/x_1)/2.$
In the Schr\"odinger equation the corresponding gauge
transformation is $\mathcal{G}=exp(iq\chi/c\hbar).$ Under this
gauge transformation the Hamiltonian $H_{\perp}(x_1,x_2)$ in
Eq.~(\ref{Eq:H1}) is transformed into
$H_{\perp} \to
\mathcal{G}H_{\perp}\mathcal{G}^{-1}=H_{\perp}^{\;\prime}
=\left(p_i + \mu\omega_c \epsilon_{ij}x_j/2\right)^2/2\mu +
\mu\omega_P^2 x_i^2/2.$
Here $H_{\perp}^{\;\prime}$ is the same $\hat H_{\perp}.$

\vspace{0.1cm}

In the limit of the kinetic energy approaching its lowest
eigenvalue the corresponding reduced constraints are the same
$\hat \varphi_i$ in Eq.~(\ref{Eq:C3}).
Under the gauge transformation $\mathcal{G}$ the angular momentum
$J_z=\epsilon_{ij}x_i p_j$ is transformed into
$J_z \to
\mathcal{G}J_z\mathcal{G}^{-1}=J_z^{\prime}=x_1p_2-x_2p_1+
q\Phi_0/2\pi c.$
Using the constraints $\hat\varphi_i$ in Eq.~(\ref{Eq:C3}) to
represent $p_1$ and $p_2$ by, respectively, the independent
variables $x_2$ and $x_1$, the first term in $J_z^{\prime}$ reads
$x_1p_2-x_2p_1=\mu\omega_c(x_1^2+x_2^2)/2.$
Thus we obtain
\begin{equation}
\label{Eq:J5}
J_z^{\prime}=\frac{q}{2\pi c}\Phi_0 +
\frac{1}{2}\mu\omega_c(x_1^2+x_2^2).
\end{equation}
$J_z^{\prime}$ is the same $J_z$ in Eq.~(\ref{Eq:J}). This result
shows that the fractional zero-point angular momentum induced by
the B-A vector potential is a real physical observable which
cannot be gauged away by a gauge transformation.

\vspace{0.1cm}

In summary, this paper explores a ``spectator" mechanism in B-A
effects. It is clarified that the B-A vector potential alone
cannot lead to non-trivial dynamics at quantum mechanical level in
the limit of the kinetic energy approaching one of its
eigenvalues. In such a limit the B-A vector potential alone cannot
induce a fractional zero-point angular momentum. When there is a
``spectator" magnetic field the B-A vector potential induces a
fractional zero-point angular momentum. The induced effect
essentially depends upon the participation of a ``spectator"
magnetic field. The ``spectator" vector potential does not
contribute to the fractional angular momentum, but plays essential
role in guaranteeing non-trivial dynamics at quantum mechanical
level in the required limit. The ``spectator" mechanism is
significant in both aspects of theory and experiment. In the
theoretical aspect, it is revealed that, unlike ordinary vector
potentials, the physical role played by the B-A vector potential
is subtle. This needs to be carefully analyzed at quantum
mechanical level. In the experimental aspect, existence of a
``spectator" magnetic field is necessary for inducing the
fractional angular momentum by the B-A vector potential. As an
example, the modified combined trap provides a realistic way to
realize this ``spectator" mechanism.

\vspace{0.4cm}

The author would like to thank M. Peshkin for the helpful
discussions and Referee for the valuable comments. This work has
been supported by the Natural Science Foundation of China under
the grant number 10575037 and by the Shanghai Education
Development Foundation.



\begin{thebibliography}{99}
\bibitem{note-1}
The derivation of a static effective potential $V_{eff}$ from a
rapidly oscillating one in Ref.~\cite{CSW} does not include a
magnetic field or an associated vector potential. That derivation
remains valid in the presence of magnetic fields, see
Ref.~\cite{BG} of the combined trap.
A modified combined trap includs a B-A type magnetic field. The
B-A effect consists essentially in a shift of the phase of the
original wave function.
%
One can adjust the radio-frequency $\tilde{\Omega}$ to compensate
the phase shift, therefore for a modified combined trap the
derivation in Ref.~\cite{CSW} also remains valid. The modification
of $\tilde{\Omega}$ leads to the corresponding modification of the
effective frequecy $\omega_P=\Omega^2/4\tilde{\Omega}$ of
$V_{eff}$. In Eq.~(\ref{Eq:H1}) the $\omega_P$ means the modified
effective frequecy.
\bibitem{constraint}
The momentum $p_{i}$ of Eq.~(\ref{Eq:p1}) defined from the
Lagrangian $L_0$ of Eq.~(\ref{Eq:L2}) cannot determine the
velocity $\dot{x_i}$ as a function of $p_{i}$ and $x_j$. This
shows that $L_0$ is singular. The physical meaning of
Eq.~(\ref{Eq:p1}) is that the corresponding Eq.~(\ref{Eq:C1}),
$\varphi_i(x,p)=0$, is a primary constraint which expresses the
dependence of degrees of freedom among $p_{i}$ and $x_j$. The
Hamiltonian equations of such a constrained system are not unique.
The standard way of deriving them is as follows.

From the Hamiltonian $H_0=p_i\dot{x_i}-L_0$, using
%
$p_{i}=\partial L_0/\partial \dot{x_i}$
%
and the Lagragian equation
$\dot{p_i} =\partial L_0/\partial x_i,$
it follows that
\begin{equation}
\label{Eq:Heq-1}
\delta H_0=\dot{x_i}\delta p_i-\dot{p_i}\delta x_i.
\nonumber
\end{equation}
It indicates that $H_0$ can be expressed as a function of $x_i$
and $p_i$.
%
%
Thus we obtain
$$\delta H_0(x,p)
=\frac{\partial H_0}{\partial x_i}\delta{x_i}
%
+\frac{\partial H_0}{\partial p_i}\delta{p_i}.$$
Because of the constraints $\varphi_i(x,p)=0$ of
Eq.~(\ref{Eq:C1}), $H_0$ plus any linear combination of
$\varphi_i$ is also a Hamiltonian of the system, i. e., the $H_0$
can be replaced by
%
$H_0(x,p)+\lambda_i(x,p)\varphi_i(x,p).$
%
From the above two equations, including the contributions of
$\delta \left(\lambda_i(x,p)\varphi_i(x,p)\right),$
it
follows that the Hamiltonian equations read
$$\dot{p_i}=-\frac{\partial H_0}{\partial
x_i}-\lambda_k\frac{\partial \varphi_k}{\partial x_i}, \;
%
\dot{x_i}=\frac{\partial H_0}{\partial
p_i}+\lambda_k\frac{\partial \varphi_k}{\partial p_i}.$$
Eq.~(\ref{Eq:C1}) gives
%
$\partial \varphi_k/\partial p_i=\delta_{ki}.$
%
From the reduced Hamitonian $H_0$ obtained from $L_0$ in
Eq.~(\ref{Eq:L2}) it follows that
$\partial H_0/\partial p_i=0.$
%
Thus the second equation reduces to
%
$$\dot{x_i}=\lambda_i.$$
%
In this example the Lagrage multipier $\lambda_i$ is just the
velocity $\dot{x_i}.$
\bibitem{note-3}
Ref.~\cite{PT60} investigated the angular momentum ${\bf J}$
originated from the Poynting vector produced by crossing the
Coulomb field ${\bf E}$ of a charged particle with an external
magnetic field ${\bf B}$,
$${\bf J}=
\frac{1}{4\pi c}\int{\bf r}\times[{\bf E}\times{\bf B(r)}]d^3 r.$$
In cases where the magnetic field is only in the $z$-direction,
this angular momentum reduces to %
$$J_z=-\frac{q\phi}{2\pi c},$$
where
$\phi=\int\int B_z(x_1,x_2)dx_1dx_2$
is the total magnetic flux. $J_z$ is the angular momentum of {\it
the electromagnetic fields}. In cases where the magnetic field
$B_z$ is produced by an infinitely long solenoid, this angular
momentum exists {\it only} inside the solenoid. $J_z$ should be
distinguished from $\mathcal{J}_{AB}$ of Eg.~(\ref{Eq:J-AB}).
$\mathcal{J}_{AB}$ is the angular momentum of {\it the charged
particle}. It is worth noting that $\mathcal{J}_{AB}$ is induced
by the B-A vector potential outside the solenoid and does {\it
not} exist inside the solenloid.
\bibitem{note-4}
This gauge function is singular at $x_1=0$. The values of the
polar angle $\phi$ on both sides of $x_2=0$, $x_1<0$ differ by
$2\pi$.  There is a cut along the negative semi-axis of $x_1$, the
so called Dirac string. However, there is no need to perform a
singular gauge transformation. See, for example,
Ref.~\cite{M-K06}.
\end{thebibliography}
\end{document}